\documentclass[%
 aip,
 amsmath,amssymb,
 reprint,%
]{revtex4-1}

\usepackage{xcolor}    
\usepackage[utf8]{inputenc}
\usepackage{graphicx}
\usepackage{dcolumn}
\usepackage{bm}
\usepackage{hyperref}
\usepackage[mathlines]{lineno}
\usepackage{amsmath} 
\usepackage{siunitx}   
\usepackage{multirow}  
\usepackage{textcomp}  

\usepackage{upgreek}

\relax 

\begin{document}

\title{Direct calibration of microwave amplification chain on an axion cavity haloscope}

\author{Hsin~Chang}\affiliation{Department of Physics, National Central University, Taoyuan City 320317, Taiwan}
\author{Han-Wen~Liu}\affiliation{Department of Physics, National Central University, Taoyuan City 320317, Taiwan}
\author{Hien~Thi~Doan}\affiliation{Institute of Physics, Academia Sinica, Taipei City 115201, Taiwan} 
\author{Yung-Fu~Chen}\email[Correspondence to: ]{yfuchen@ncu.edu.tw}\affiliation{Department of Physics, National Central University, Taoyuan City 320317, Taiwan}

\date{\today}

\begin{abstract}  

In an axion haloscope, the weak photon signal, theoretically converted from axions, is captured by a detection cavity. The signal from the cavity is too weak to be acquired by a signal receiver. The amplification chain assists the signal acquisition by amplifying the signal and requires accurate gain calibration. Typically, the readout line is calibrated using the Y-factor method, involving a switch that directs either the detection line or the calibration line to the amplification chain. The detection and calibration lines may have different transmissions, which leads to the calibration results being less accurate. In response, we propose a calibration method that eliminates the need for a switch. In this approach, the cavity temperature is decoupled to its incoming noise source and can be controlled, resulting in excess or deficiency of the noise spectrum near its resonance frequency. The experimental result shows that the gain of the amplification chain can be calibrated directly using the temperature-varied cavity radiation.

\end{abstract}

\date{\today}

\maketitle

\section{Introduction}

In an axion~\cite{peccei1977cp,peccei1977constraints,weinberg1978new,wilczek1978problem} haloscope, a cavity immersed in a strong magnetic field is used as a detector~\cite{sikivie1983experimental,sikivie1985detection,sikivie2021invisible}. The photons converted from the axions are stored in the cavity when the cavity resonance frequency $f_{\rm c}$ matches the photon frequency $f_{\rm a} \equiv m_{\rm a}c^2/h$, where $c$ is the speed of light, $m_{\rm a}$ is the mass of axion, and $h$ is the Planck constant. The photons in the cavity are captured via an antenna, amplified by an amplification chain, and measured by a signal receiver. Besides the axion-converted photons, the cavity emits quantum or thermal noise since it is thermalized to its environment. The temperature of the cavity $T_{\rm c}$ should be kept at $T_{\rm c} \ll hf_{\rm c}/k_{\rm B}$ to reduce the thermal noise in measurement, where $k_{\rm B}$ is Boltzmann constant. Generally, a dilution refrigerator is used to host the cavity~\cite{chang2022first,chang2022taiwanaxion,al2017design,khatiwada2021axion,choi2021capp} as it provides a continuous cooling for massive materials down to the order of 10 mK. Moreover, the axion-to-photon converted signal power is extremely weak. For most advanced haloscope setups, the power is yet below $10^{-22}\ \rm{W}$~\cite{borsanyi2016calculation,dine2017axions,hiramatsu2011improved,kawasaki2015axion,berkowitz2015lattice,petreczky2016topological,ballesteros2017unifying,buschmann2020early,hagmann1998results,asztalos2010squid,bartram2021search,brubaker2017first,zhong2018results,backes2021quantum,kwon2021first}. To measure the conversion signals, linear amplification is often preferred~\cite{hagmann1998results,brubaker2017first,kwon2021first,chang2022first,lamoreaux2013analysis}, while other methods, such as single-photon detection~\cite{pankratov2024observation,braggio2024quantum,kuo2024maximizing}, are also under exploration. The detection sensitivity can be determined once the system noise and the predicted signal power are known. To obtain the system noise, calibrating the gain of the amplification chain is required. The added noise of the amplification chain can also be known to indicate the performance of the system.

\begin{figure}[btph]
    \centering
    \includegraphics[width=0.48\textwidth]{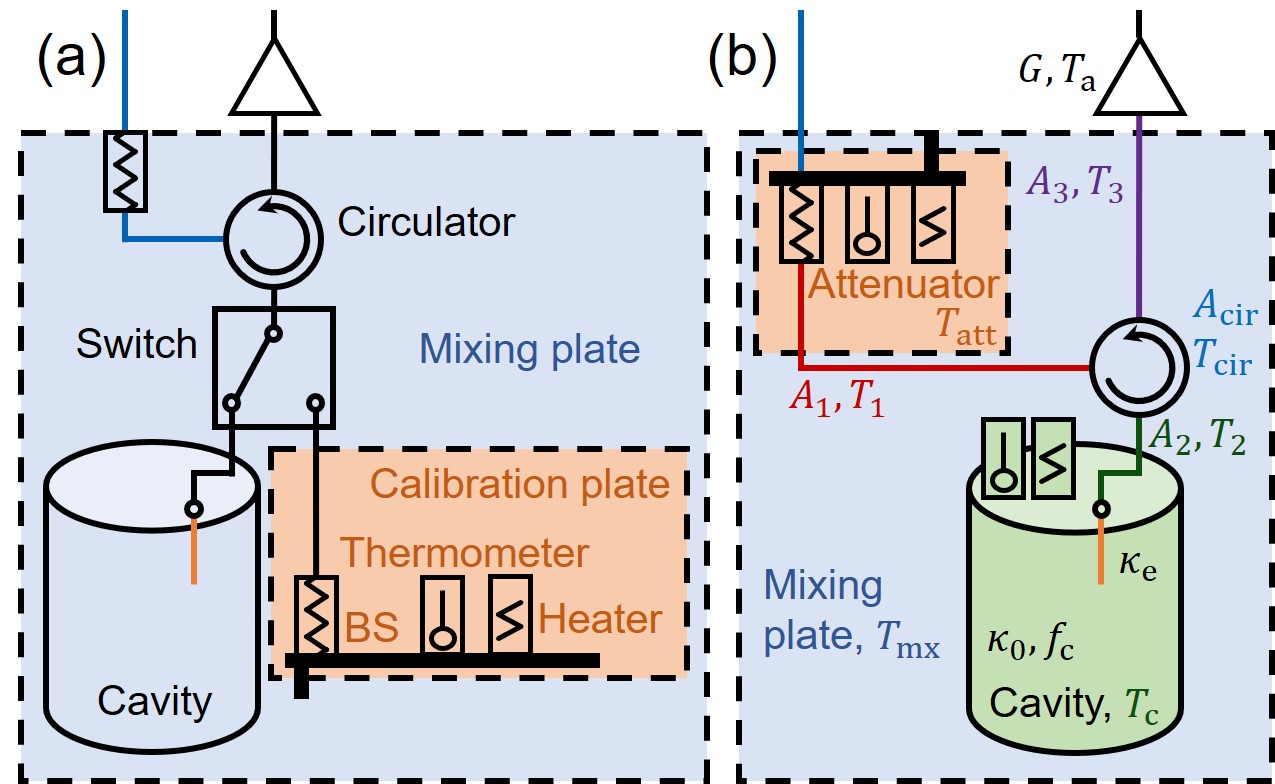}
    \caption{Axion haloscope setup. (a) Y-factor calibration setup. A cryogenic switch connects the readout chain to the cavity or the black-body radiation source. (b) Cavity radiation calibration setup. The amplification chain is calibrated directly by cavity radiation.}
    \label{fig:setup}
\end{figure}

To calibrate the linear amplification chain, a source that emits a known and adjustable power $P_{\rm s}$ is required. During the calibration process, the power measured at the signal receiver $P_{\rm r}$ depends on $P_{\rm s}$ linearly. The slope and the intersection with the y-axis of $P_{\rm r}$ versus $P_{\rm s}$ give the gain and added noise of the amplification chain, respectively. Typically, a Y-factor method, as illustrated in Fig. \ref{fig:setup}(a), is used~\cite{chang2022taiwanaxion,alesini2019galactic,lee2022searching}. For gain calibration, a switch directs the readout to the black-body radiation source (BS), whose emitted power is adjusted and measured by the heater and thermometer, respectively. The BS serves as the source of $P_{\rm s}$ and is used to calibrate the gain from the BS to the signal receiver. For axion detection, the switch connects the amplification chain to the cavity. An antenna (marked in orange in Fig. 1) is inserted into a microwave (MW) cavity as a probe. The emitted power at the cavity is inferred via the calibration. However, the detection and calibration have different transmission paths before the switch. The different attenuation of cables between the paths leads to the calibration results being less accurate. Besides, the insertion loss of the switch may not be consistent in every operation.

To avoid the need for switches, different calibration methods have been developed~\cite{braggio2022haloscope,crescini2020axion}. Here, we propose using the cavity as a calibration source so that the detection path is directly calibrated. Since no switch is involved, the problems in the Y-factor method are mitigated. The gain obtained in the calibration process can be applied to the axion search without the concerns mentioned above. A model is introduced to derive the noise spectrum from a cavity that is not thermalized with its incoming noise source. The losses of MW components are considered in the model to make the calibration results more accurate. The model agreed with the experimental data from a hot cavity at resonance frequency $f_{\rm c} \approx$ 5 GHz. The result indicates that a hot cavity can manifest its temperature through its noise spectrum near the resonance frequency, and cavity radiation is a reliable noise source for amplification chain calibration.

\section{Cavity radiation calibration}\label{Sec:Model}

Past experiments have indicated that a hot cavity emits an excess noise power near its resonance~\cite{al2017design,chang2022taiwan}. The width of the excess power is about the cavity bandwidth. The phenomenon was explained as follows: The entire noise spectrum is determined by the incoming noise reflected by the cavity and the thermal noise emitted by the cavity at an elevated temperature~\cite{chang2022taiwan,alesini2021search,wurtz2021cavity}. The previous studies on cavity radiation inspire us to use the cavity as the calibration source to avoid the disadvantages of the Y-factor method. As depicted in Fig. \ref{fig:setup}(b), the cavity is inserted with an antenna and connected to an attenuator as an incoming noise source. The temperature of the attenuator and the cavity are denoted as $T_{\rm att}$ and $T_{\rm c}$, respectively. The circulator allows the measurement of the reflection scattering parameter (S-parameter) by injecting a continuous wave from the blue cable. The scattering characteristics of the cavity are governed by intrinsic loss rate $\kappa_0$, external coupling rate $\kappa_{\rm e}$ to the antenna, and the cavity resonance frequency $f_{\rm c}$. Cable 1, 2, and 3 are marked in red, green, and purple with loss $A_1$, $A_2$, and $A_3$, respectively. The cavity radiation is added a noise $T_{\rm a}$ and amplified with gain $G$ via the amplifier. To simplify the discussion, we first focus on the cavity radiation spectrum under the assumption of cables with no loss, i.e., $A_1 = A_2 = A_3 = 1$, highlighting its potential as a calibration source. Then, the effect of cable loss is investigated to provide a more accurate representation of the practical setup. Moreover, since the cables at both ends contact with different temperatures, a linear temperature distribution of the cables, $T_1$, $T_2$, and $T_3$, has also been considered.

As indicated in Fig. \ref{fig:setup}(b), the cavity radiation consists of noise from the attenuator and the cavity. Thus, the cavity radiation power under the condition of the lossless cables is given by~\cite{chang2022taiwan,wurtz2021cavity}
\begin{equation}
\label{eq:CavityEmitted}
P_{\rm c}(f)
= \left(|S_{\rm r}|^2\widetilde{T}_{\rm att} + |S_{\rm t}|^2\widetilde{T}_{\rm c}\right)k_{\rm B}B,
\end{equation}
where
\begin{subequations}
\begin{align}
    S_{\rm r}(f) &= \frac{\kappa_0-\kappa_{\rm e} + 2i\Delta}{\kappa_0+\kappa_{\rm e} + 2i\Delta},\label{eq:Sr}\\
    S_{\rm t}(f) &= \frac{2\sqrt{\kappa_0\kappa_{\rm e}}}{\kappa_0+\kappa_{\rm e}+2i\Delta}\label{eq:St}
\end{align}
\end{subequations}
is the reflection factor of the antenna and the transmission factor from the cavity to the antenna, respectively, $B$ is the frequency resolution bandwidth of spectrum analysis, $\Delta = f-f_{\rm c}$ is the frequency detuning from $f_{\rm c}$, and $f$ is the frequency. The effective noise temperature $\widetilde{T}_{\rm x} = \widetilde{T}(T_{\rm x})$ is expressed as 
\begin{equation}
\label{eq:ENT}
    \widetilde{T}(T_{\rm x}) = \frac{hf}{k_{\rm B}} \left( \frac{1}{e^{hf/k_{\rm B}T_{\rm x}} - 1} +\frac{1}{2}\right)
\end{equation}
for the physical temperature $T_{\rm x}$, $\rm x$ stands for various physical objects. Because $\kappa_0$, $\kappa_{\rm e}$, and $f_{\rm c}$ can be obtained from the reflection S-parameter measurement, $|S_{\rm r}|^2$ and $|S_{\rm t}|^2$ in Eq. \ref{eq:CavityEmitted} can be calculated. $\widetilde{T}_{\rm att}$ and $\widetilde{T}_{\rm c}$ are determined by Eq. \ref{eq:ENT}. The known and adjustable $P_{\rm c}$ provides source power, $P_{\rm s}$, and the power after amplification is
\begin{equation}
\label{eq:CavityEmittedCal}
    P_{\rm r}(f) = G\left( P_{\rm c} + k_{\rm B}T_{\rm a}B \right).
\end{equation}
The spectrum of $P_{\rm r}$ follows that of $P_{\rm c}$, and exhibits a peak near $f_{\rm c}$ when $T_{\rm att} < T_{\rm c}$, and a dip when $T_{\rm att} > T_{\rm c}$. When $T_{\rm att} = T_{\rm c}$, it corresponds to white noise since $|S_{\rm r}|^2 + |S_{\rm t}|^2 = 1$.

The discussion up to this point has considered lossless MW components, whereas cables and circulators typically exhibit attenuation. We first look into the effect of loss on the circulator. Consider the noise $\widetilde{T}_{\rm att}$ transmitted through the circulator with a loss $A_{\rm cir}$ and temperature $T_{\rm cir}$. Based on the fluctuation-dissipation theorem, the output consists of the attenuated $\widetilde{T}_{\rm att}$ and the additional noise introduced by the circulator, as described by $A_{\rm cir}\widetilde{T}_{\rm att} + (1 - A_{\rm cir}) \widetilde{T}_{\rm cir}$. For the noise generated by cable 1, the cable temperature as a function of space, $T_1 = T_1(x)$, should be considered. The spatial coordinate of the cable $x$ is defined from the attenuator $x_{\rm att}$ to the circulator $x_{\rm cir}$, with corresponding temperature $T_{\rm att}$ and $T_{\rm cir}$. The generated noise of the cable is given as the integral of effective noise temperature at $x$ attenuated by the cable with length $|x_{\rm cir}-x|$ with respect to $x$, that is
\begin{equation}
\label{eq:con_noise}
    \widetilde{T}_1^* = \int_{x_{\rm att}}^{x_{\rm cir}} \widetilde{T}_1(x)A_1(x)\frac{dx}{l},
\end{equation}
where $\widetilde{T}_1(x) = \widetilde{T}\left[T_1(x)\right]$, $T_1(x) = T_{\rm att} + \frac{x-x_{\rm att}}{x_{\rm cir}-x_{\rm att}} (T_{\rm cir}-T_{\rm att})$ due to the assumption of constant thermal conductivity in the cable, $A_1(x) = e^{-|x_{\rm cir}-x|/l}$, and $l$ represents the attenuation length. The total loss of cable 1 is expressed as $A_1 = e^{-|x_{\rm cir}-x_{\rm att}|/l}$. This applies to all cables.

\begin{figure}[bpht]
    \centering
    \includegraphics[width=0.48\textwidth]{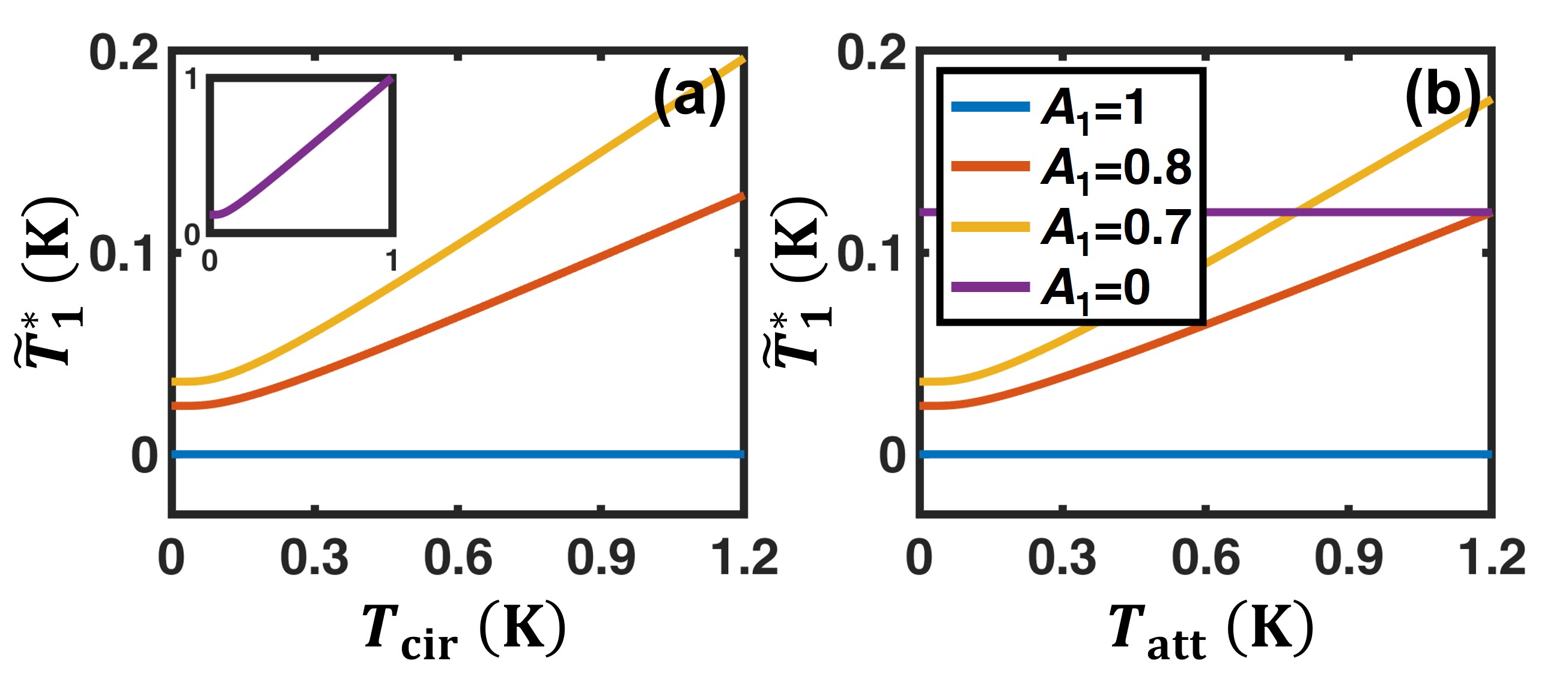}
    \caption{The generated noise from the cable $\widetilde{T}_1^*$. The plot is based on Eq. \ref{eq:con_noise} where $f = 5\ {\rm GHz}$, and $A_1$ is set at 1, 0.8, 0.7, and 0. (a) $\widetilde{T}_1^*$ as a function of $T_{\rm cir}$ where $T_{\rm att}$ is fixed at $12\ {\rm mK}$. (b) $\widetilde{T}_1^*$ as a function of $T_{\rm att}$ for $T_{\rm cir} = 12\ {\rm mK}$.}
    \label{fig:CabNoise}
\end{figure}

Figure \ref{fig:CabNoise} shows the integral of Eq. \ref{eq:con_noise} with $f = 5\ {\rm GHz}$ at $A_1 = 1$, 0.8, 0.7, and 0. The y-axis in (a) and (b) are set in the same range for comparison. Figure \ref{fig:CabNoise}(a) illustrates $\widetilde{T}_1^*$ as a function of $T_{\rm cir}$ for $T_{\rm att} = 12\ {\rm mK}$, which is much lower than cross-over temperature $hf/2k_{\rm B} = 120\ {\rm mK}$. A lossless cable, as $A_1 = 1$ shown by the blue line, does not generate noise. As the loss increases (smaller $A_1$), $\widetilde{T}_1^*$ increases. The inset shows that the cable becomes a thermal insulator when $A_1 = 0$, and $\widetilde{T}_1^*$ is independent of $T_{\rm att}$ and equals $\widetilde{T}_{\rm cir}$. Figure \ref{fig:CabNoise}(b) depicts $\widetilde{T}_1^*$ as a function of $T_{\rm att}$ with the condition of $T_{\rm cir} = 12\ {\rm mK}$. Since $A_1(x)$ assigns a weight to each $\widetilde{T}_1(x)$, $\widetilde{T}_1^*$ is more relevant to $T_{\rm cir}$. It should be noted that the generated noise depends on the direction of the signal path.

Considering the losses of the MW components, the incoming noise to the cavity should include not only the noise from the attenuator but also the generated noises from cable 1, circulator, and cable 2, and can be expressed as:
\begin{equation}
\label{eq:Tin}
    T_{\rm in} = A_2\left[ A_{\rm cir}\left( A_1\widetilde{T}_{\rm att} + \widetilde{T}_1^* \right)+(1-A_{\rm cir})\widetilde{T}_{\rm cir} \right]+\widetilde{T}_2^*.
\end{equation}
By collecting the losses, $A_{\rm in} = A_2A_{\rm cir}A_1$, and the generated noises of MW components, $\hat{T}_{\rm in} = A_2A_{\rm cir}\widetilde{T}_1^*+A_2(1-A_{\rm cir})\widetilde{T}_{\rm cir}+\widetilde{T}_2^*$, $T_{\rm in}$ is rearranged to
\begin{equation}
\label{eq:Tin_simple}
    T_{\rm in} = A_{\rm in}\widetilde{T}_{\rm att} + \hat{T}_{\rm in}.
\end{equation}
The cavity radiation power is modified as
\begin{equation}
\label{eq:CavityEmittedModified}
\begin{aligned}
    P_{\rm c}(f) = \left( |S_{\rm r}|^2T_{\rm in} + |S_{\rm t}|^2\widetilde{T}_{\rm c}\right) k_{\rm B}B.
\end{aligned}
\end{equation}
For the calibration, cable 2, circulator, and cable 3 should be included in the amplification chain. The losses and generated noises of MW components before the first-stage amplifier are collected into $A_{\rm r}=A_3A_{\rm cir}A_2$ and $\hat{T}_{\rm r} = A_3A_{\rm cir}\widetilde{T}_2^{*-}+A_3(1-A_{\rm cir})\widetilde{T}_{\rm cir}+\widetilde{T}_3^*$. $\widetilde{T}_2^{*-}$ denotes the noise generated by cable 2 in the direction outward from the cavity. The power after amplification is
\begin{equation}
\label{eq:lossCal}
\begin{aligned}
    P_{\rm r}(f) &= G\left( A_{\rm r}P_{\rm c} + k_{\rm B}\hat{T}_{\rm r}B + k_{\rm B}T_{\rm a}B \right)\\
    &= G' \left( P_{\rm c} + k_{\rm B}T_{\rm a}'B \right).
\end{aligned}
\end{equation}
An effective gain, $G' = GA_{\rm r}$, and effective added noise, $T_{\rm a}' = (\hat{T}_{\rm r} + T_{\rm a})/A_{\rm r}$, reveals that the signal is degraded by a factor of $A_{\rm r}$, while the noise is increased by a factor of $1/A_{\rm r}$. In cavity radiation calibration, $P_{\rm c}$ acts as the calibration source $P_{\rm s}$. The calibration gives $G'$, which is used for sensitivity determination and noise performance during axion search. The experimental results show the added noise included $\hat{T}_{\rm r}$ can be represented by one fitting parameter $T_{a}'$.

\section{Experimental setup} \label{sec:ExpSetup}

The setup of cavity radiation calibration is demonstrated in Fig. \ref{fig:setup}(b), which comprises the attenuator plate, marked in orange, and the cavity. The attenuator plate and the cavity are thermal-anchored via two stainless steel pillars to the mixing plate. The poor thermal conductance of pillars ensures that the temperature of the mixing plate will not be significantly affected during the heating process. Thus, $T_{\rm att}$ and $T_{\rm c}$ can be controlled independently by the resistive heaters and calibrated $\rm RuO_2$ thermometers. A good thermal anchor between the attenuator and the attenuator plate ensures that the temperature reading from the thermometer accurately expresses the temperature of the attenuator~\cite{chang2022taiwanaxion}. The cavity is a cylinder made of oxygen-free high-conductivity (OFHC) copper with a height of 57 mm, an outer diameter of 58 mm, a thickness of 5 mm, and a mass of 600 g. The cavity has a nearly uniform internal temperature distribution due to its high thermal conductivity and small size. The thermometer reading is stabilized within 10 minutes after the heater output power change. To ensure that the temperature of the attenuator and that of the cavity reach a steady state, we wait 30 minutes after each temperature adjustment.

The attenuator is not only used as a radiation source but also reduces thermal noise from the input line (cable in blue). This ensures that the incoming noise to the cavity is mainly contributed by the attenuator. Cable 1 and 2 are CryoCoax-BCB012 superconducting cables with 40 cm and 53.5 cm long, respectively, to reduce the attenuation on $\widetilde{T}_{\rm att}$. The LNF-CIISISC4\_8A circulator propagates the outgoing field from the cavity to the amplification chain and prevents thermal radiation from the amplifier from being transmitted back to the cavity. The circulator also allows reflection spectroscopy measurement. The amplification chain is composed of an LNF-LNC4\_8C low-noise high-electron-mobility-transistor (HEMT) amplifier on the 4K plate and a three-stage Mini-Circuits ZX60-83LN-S+ room-temperature post-amplifier (not shown in the figure). The power is analyzed through the fast Fourier transformation with bandwidth $B = 1\ {\rm kHz}$ and a three-minute integration time by the NI PXIe-5644R vector signal transceiver (VST).

\section{Experimental results} 

To demonstrate our calibration method, the cavity parameters are first obtained through the reflection spectroscopy measurement. Then, the cavity radiation spectrum is taken under various $T_{\rm att}$ and $T_{\rm c}$. An asymmetric spectrum for $P_{\rm r}$ is observed in our setup, presumably due to the nonuniform frequency response of the readout chain~\cite{asztalos2001large,lee2020axion}. Typically, this nonuniform frequency response in haloscope experiments is removed by digital filters, such as the Savitzky-Golay (SG) filter or Pad\'e filter, when processing the data~\cite{chang2022taiwan,brubaker2017haystac,bartram2021axion}. In this work, the frequency dependence is distinguished from the fitting of Eq. \ref{eq:lossCal}. Theoretically, both $T_{\rm a}'$ and $G'$ as a function of frequency lead to a nonuniform behavior. However, $T_{\rm a}'$ involves fewer MW components and typically behaves as a constant within a sufficiently small measurement span. Therefore, we take frequency-dependent gain $G' = G'(f)$ into account. To determine the frequency dependence, the fitting is first performed with five parameters: $\Bar{G}'$, $T_{\rm a}'$, $A_1$, $A_{\rm cir}$, and $A_2$, where $\Bar{G}'$ is assumed to be frequency independent. The fitting result of $\Bar{G}'$ corresponds to the averaged gain in the measurement frequency span. The frequency dependence is derived from the ratio between the data and the fitting curve.

\subsection{Reflection spectroscopy measurement} \label{sec:CavScaMea}

\begin{figure}[btph]
    \centering
    \includegraphics[width=0.36\textwidth]{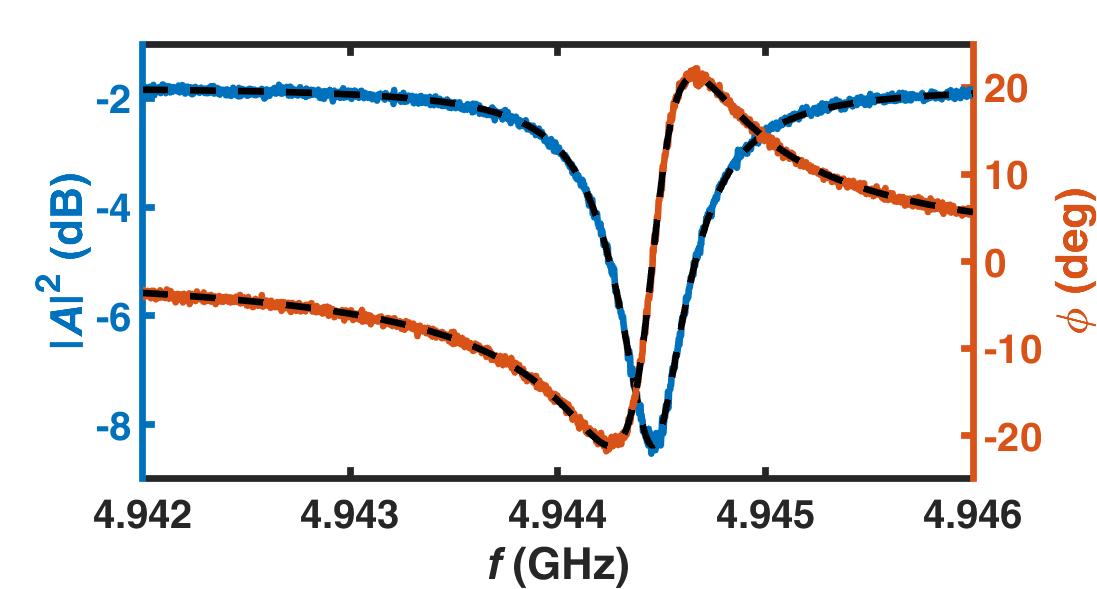}
    \caption{The cavity reflection scattering measurement. The solid lines are the experimental data. The dashed line is the fitting with Eq. \ref{eq:Sr} and gives $\kappa_0 = 431\ {\rm kHz}$, $\kappa_{\rm e} = 156\ {\rm kHz}$ and $f_{\rm c} = 4.944 455\ {\rm GHz}$.}
    \label{fig:calibration}
\end{figure}

The factors $|S_{\rm r}|^2$ and $|S_{\rm t}|^2$ in Eq. \ref{eq:CavityEmittedModified} are related to the cavity characteristics: $\kappa_0$, $\kappa_{\rm e}$, and $f_{\rm c}$, which can be determined from the reflection spectroscopy measurement. Figure \ref{fig:calibration} shows the amplitude $|A|^2$ and phase $\phi$ of the reflection spectroscopy of the cavity coupled with the external antenna. By fitting $|A|^2$ and $\phi$ using Eq. \ref{eq:Sr}, we obtained $\kappa_0 = 431 \pm 7\ {\rm kHz}$, $\kappa_{\rm e} = 156 \pm <1\ {\rm kHz}$ and $f_{\rm c} = 4.944455\ {\rm GHz} \pm 3\ {\rm kHz}$. Since far-detuning approximation brings $|S_{\rm r}|^2$ to 1, the baseline offset of $|A|^2$ is due to the attenuation of the input line and amplification of the readout line.

\subsection{Cavity radiation}

\begin{figure*}[btph]
    \centering
    \includegraphics[width=1\textwidth]{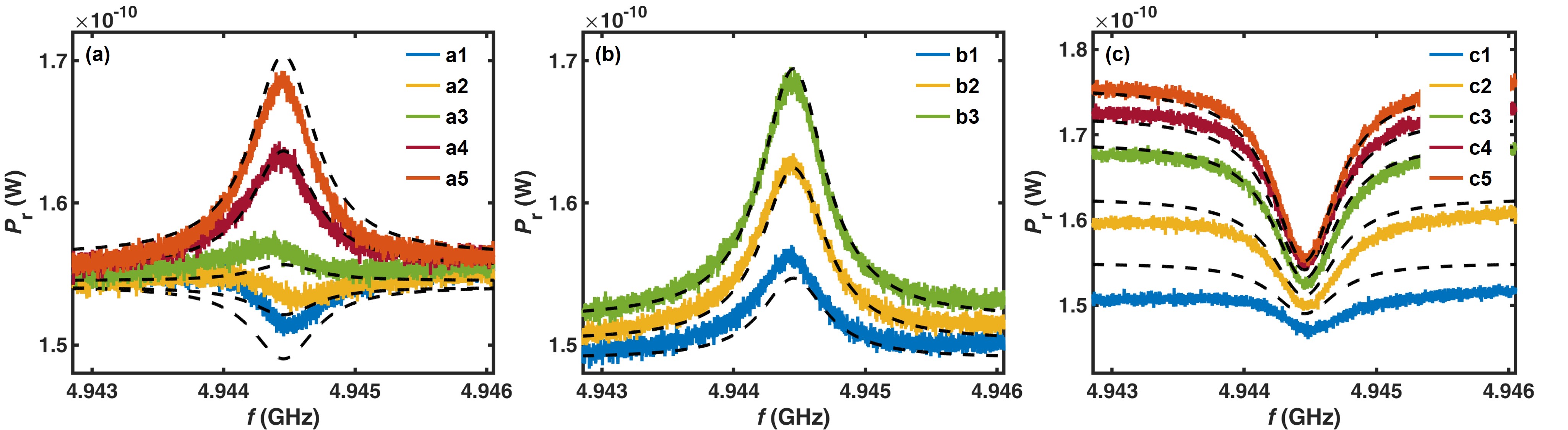}
    \caption{The cavity radiation spectrum measured by the signal receiver. The monitored temperatures are depicted in Tab. \ref{tab:compare}. The solid curves are the measurement, and the dashed curves are the fitting to Eq. \ref{eq:lossCal} with common frequency-independent gain. (a) The data collection with $T_{\rm att} \approx 340\ {\rm mK}$, and $T_{c}$ is adjusted from below to above $T_{\rm att}$. (b) The data collection with low $T_{\rm att}$ and heated $T_{\rm c}$. (c) The data collection with low $T_{\rm c}$ and heated $T_{\rm att}$.}
    \label{fig:raw}
\end{figure*}

\begin{figure*}[tbph]
    \centering
    \includegraphics[width=1.00\textwidth] {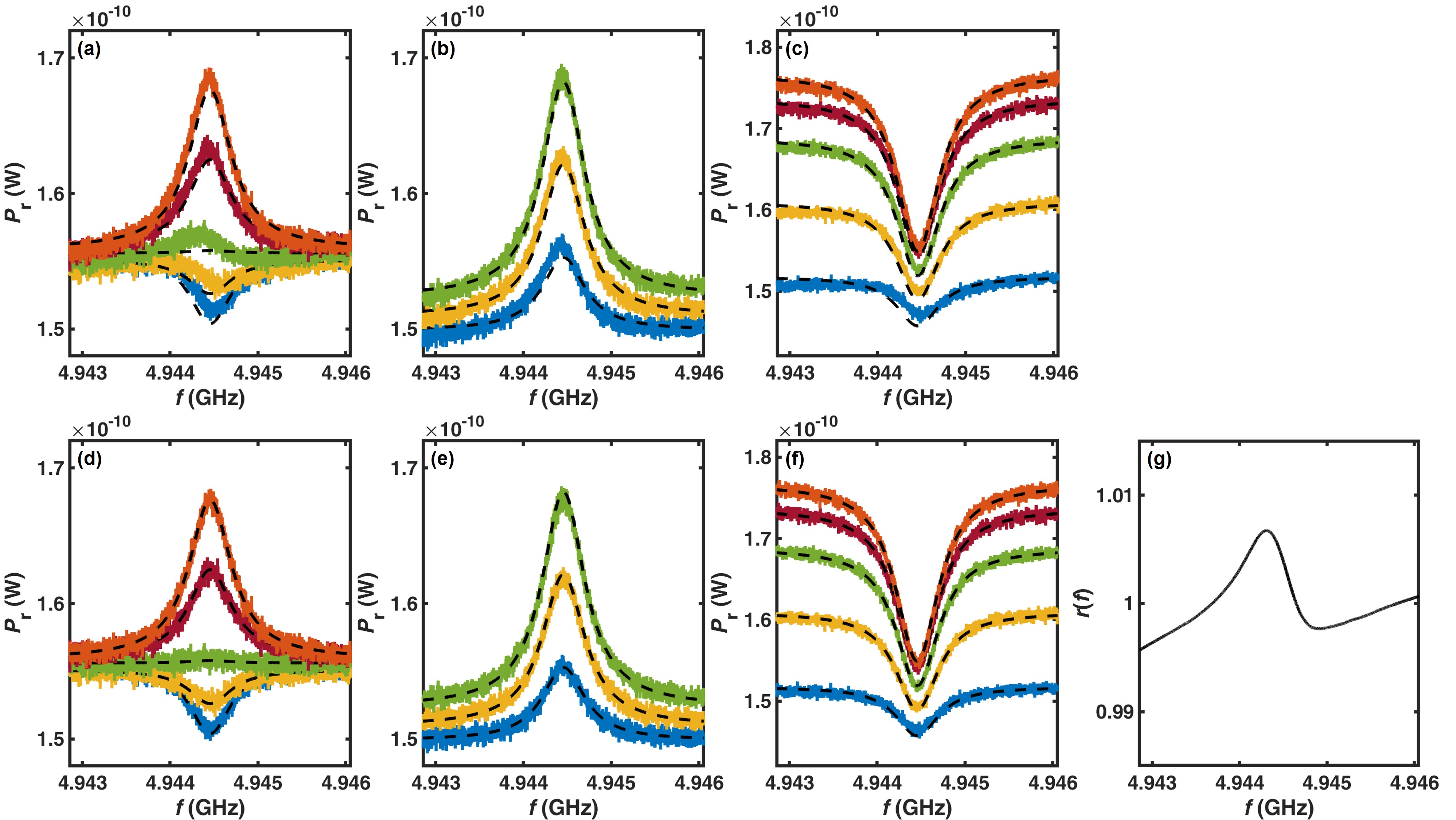} 
    \caption{The fitting models with gain variation and frequency dependence. (a)-(c) Fitting results with gain are considered to vary over datasets and frequency independent. (d)-(g) Fitting results with frequency-dependent varying gain. $r(f)$ is determined from the data fitting. The data in Fig. (a)-(c) after normalized with respect to $r(f)$ are depicted by solid curves. The dashed curves are the fitting to Eq. \ref{eq:lossCal}.}
    \label{fig:rf}
\end{figure*}

\begin{table}[b]
    \centering
    \begin{tabular}{|c|c|c|c|c|l|c|c|c|c|c|}
    \hline
         Curve & $T_{\rm att}$ & $T_{\rm c}$ & $T_{\rm mx}$ & $\Bar{G}'$ & \multicolumn{2}{|c|}{Curve} & $T_{\rm att}$ & $T_{\rm c}$ & $T_{\rm mx}$ & $\Bar{G}'$ \\ \hline
         a1& 347& 78& 23& 92.91& \multicolumn{2}{|c|}{c1}& 384& 70& 23& 92.77 \\ \hline
         a2& 335& 202& 25& 92.90& \multicolumn{2}{|c|}{c2}& 710& 88& 32& 92.80 \\ \hline 
         a3& 340& 338& 30& 92.91& \multicolumn{2}{|c|}{c3}& 985& 110& 42& 92.84 \\ \hline 
         a4& 344& 648& 46& 92.91& \multicolumn{2}{|c|}{c4}& 1116& 124& 48& 92.88 \\ \hline
         a5& 345& 909& 63& 92.90& \multicolumn{2}{|c|}{c5}& 1257& 136& 54& 92.86 \\ \hline
         b1& 86& 348& 28& 92.93& \multicolumn{2}{|c|}{}&&&& \\ \hline
         b2& 104& 646& 44& 92.94& \multicolumn{2}{|c|}{}&&&& \\ \hline
         b3& 143& 910& 63& 92.95& \multicolumn{2}{|c|}{}&&&& \\ \hline
         \end{tabular}
\caption{The readings of thermometers on the attenuator plate, the cavity, and the mixing plate during the data collection in Fig. \ref{fig:raw} and the related fitting results for the gain-drifting model. The temperatures are in units of ${\rm mK}$, and gains are in units of ${\rm dB}$.}
\label{tab:compare}
\end{table}

The solid curves in Fig. \ref{fig:raw} show the measured cavity radiation spectrum under various combinations of $T_{\rm att}$ and $T_{\rm c}$. The asymmetric shapes are caused by $G'(f)$. The temperatures monitored by thermometers are listed in Tab. \ref{tab:compare}. $T_{\rm cir} = T_{\rm mx}$ is assumed because the circulator attaches to the mixing plate. We first fit data with frequency-independent gain to Eq. \ref{eq:lossCal}. We obtain $\Bar{G}' = 93.68 \pm \ 0.14{\rm dB}$, $T_{\rm a}' = 4.50 \pm 0.22\ {\rm K}$, $A_1 = 0.58 \pm 0.08$, $A_{\rm cir} = 1.00 \pm 0.02$, and $A_2 = 0.87 \pm 0.06$, as shown in Fig. \ref{fig:raw}. The reduced chi-square is $\chi^2_{\upnu} = 14.34$. The large $\chi^2_{\upnu}$ may be due to the lack of consideration for the gain drifting over time, the transmissivity change of the other nearby microwave components due to heating during calibration, and the gain frequency dependence. Considering the gain variation over different datasets, the thirteen datasets are fitted with an individual gain but share a common added noise and attenuation of the MW components. We obtain $\langle\Bar{G}'\rangle = 92.89 \pm 0.05\ {\rm dB}$, which is the average gain across the thirteen fitting results, $T_{\rm a}' = 5.44 \pm 0.07\ {\rm K}$, $A_1 = 0.98 \pm 0.02$, $A_{\rm cir} = 0.98 \pm 0.01$, and $A_2 = 0.96 \pm 0.02$. The fitting curves are shown in Fig. \ref{fig:rf} (a)-(c) with $\chi^2_{\upnu} = 2.58$. We first verify the fitting results by comparing the estimated losses to the values from the specifications provided by manufacturers. As the instrument models introduced in Sec. \ref{sec:ExpSetup}, the losses of the cables and the insertion loss of the circulator are $A_1 = 0.96$, $A_2 = 0.94$, and $A_{\rm cir} = 0.96$, which are close to the fitting results. The individual fitting results for $\Bar{G}'$ are listed in Tab. \ref{tab:compare} and show a $\sigma_{\rm \Bar{G}'}/\langle\Bar{G}'\rangle = 1.21\%$ varying over a nine-hour experimental time, where $\sigma_{\rm \Bar{G}'}$ is the standard deviation of $\Bar{G}'$. The gain drifting may be attributed to the variation in the VST module temperature~\cite{chang2022taiwanaxion} or the drifting gain of the amplifiers.

The discrepancy between data and the fitting curves in Fig. \ref{fig:rf}(a)-(c) is assumed to be due to the frequency dependence of the readout line gain, which is expressed as $G'(f) = r(f)\Bar{G}'$. The ratio $r(f)$ is determined by averaging the ratio between the data and the fitting of thirteen data curves and smoothing the averaged result via the SG filter. Figure \ref{fig:rf}(g) plots $r(f)$ and reveals that the gain variation in frequency is $0.55\%$. Note that the structure of r(f) near cavity resonance frequency is likely due to the Fano resonance effect~\cite{rieger2023fano}. The effect should be quite weak and not easy to disentangle from the frequency dependence of other components in the readout line. Therefore, We lumped all effects into the frequency dependence gain. The data normalized by $r(f)$ are plotted in Fig. \ref{fig:rf}(d)-(f), yielding $\chi^2_{\upnu} = 1.22$. The number of points to describe $r(f)$ is considered in the degree of freedom. The reduction in $\chi^2_{\upnu}$ value indicates the necessity of considering gain variations over time and frequency. The $\chi^2_{\upnu}$ slightly larger than 1 may result from systematic errors introduced by the lack of consideration for a drifting $r(f)$ or the inaccurate assumption for temperature distribution of the cables, but the effect is small enough to give a good approximation to determine $\Bar{G}'$. We test the model with the assumption that the thermal conductivity of the cables is linearly proportional to temperature. It shows that $\langle \Bar{G}' \rangle$ differs 0.62\% from the aforementioned fitting result, where the thermal conductivity of cables is assumed to be constant. Because the fitting result shows that the cable losses are close to 1, the noise generated by the cable makes a small contribution to the cavity radiation power and does not significantly affect the calibrated gain.  

\section{Applicability}

In this section, we explore the applicability of cavity radiation calibration in practical axion search experiments. The duty cycle of the calibration method is influenced by the heat capacity and thermal conductivity of the cavity, as well as the cooling power of the cold plate in contact with the cavity. In the setup of this experiment, the cavity, as described in Sec. \ref{sec:ExpSetup}, requires 30 minutes to return to the base temperature. In the high-frequency axion-searching range, a small size cavity~\cite{chang2022taiwanaxion} allows its temperature to change easily, making cavity radiation calibration by adjusting cavity temperature more feasible. However, to compensate for the volume loss, the large cavity design for high-frequency searching, such as a multiple-cell cavity~\cite{jeong2020search}, is considered. A large cavity is also used in the low-frequency searching range~\cite{khatiwada2021axion}. A copper-plated stainless-steel cavity is considered to enhance the mechanical strength of the cavity~\cite{boutan2024axion}. For a cavity with large size or low thermal conductivity, it takes a longer time to adjust the cavity temperature. Cavity radiation calibration can instead be performed by adjusting the incoming noise source temperature, which is $T_{\rm att}$ in this experiment~\cite{simbierowicz2021characterizing}. Performing the calibration by changing cavity temperature can be considered supplementary.

To enhance the detection performance of the haloscope setup, a parametric amplifier (PA) is usually introduced as the first-stage amplifier in the amplification chain. PA needs to be operated at sub-Kelvin temperatures and is usually installed on the mixing plate. Due to the proximity of the PA to the cavity, heating the cavity could impact the PA's gain performance. The cavity radiation calibration cannot be applied directly. However, the MW propagation in a PA chip does not change with temperature up to $\sim1\ {\rm K}$ when PA is off. Because the PA fully reflects the input signal even at elevated temperatures when its pump is off, cavity radiation calibration is feasible for calibrating the amplification chain with the pump-off PA. The PA gain can be independently calibrated using a pump on/off method~\cite{braggio2022haloscope}. Combining the two calibration results gives the calibrated gain performance when the PA pump is on.

\section{Conclusion}

In this work, We modify the haloscope setup by directly using the cavity as the calibration source, enabling the calibration and detection to share the same path. The absence of a switch eliminates concerns of different attenuation among coaxial cables, heating, and inconsistent connection from the operating switch. Additionally, the frequency dependence of the readout line gain can be identified by the analysis of the measured cavity radiation spectrum. Furthermore, the impact of lossy cables has been studied, and the experimental results are in agreement. The consideration of losses for the incoming noise offers a more accurate description of the realistic behavior of the cavity-emitted spectrum. The cavity radiation method has been properly understood and offers a broadly applicable and more reliable calibration process across various experimental setups.

\begin{acknowledgments}
We are grateful to Yuan-Hann~Chang for their helpful discussions. The work is strongly supported by the Taiwan Axion Search Experiment with Haloscope (TASEH) collaboration. The work was supported by the National Science and Technology Council in Taiwan under grant No. NSTC 111-2123-M-001-004 and No. NSTC 112-2123-M-001-001.
\end{acknowledgments}

\section*{Data availability}
The data that support the findings of this study are available from the corresponding author upon reasonable request.

\end{document}